\documentclass[12pt]{iopart}

\usepackage{graphicx,color} 

\usepackage[latin1]{inputenc}
\usepackage[OT1]{fontenc}
\usepackage[english]{babel} 
\usepackage{booktabs,multirow}
\usepackage{float} 

\usepackage{harvard}

\begin{document}






\title[A Novel Wavelet-based Filtering Strategy to Remove PLI from ECGs]{\Large {A Novel Wavelet-based Filtering Strategy to Remove Powerline Interference from Electrocardiograms with Atrial Fibrillation}}

\author{Manuel García$^{1}$, Miguel Martínez-Iniesta$^{1}$, Juan Ródenas$^1$, José J. Rieta$^2$ and Raúl Alcaraz$^1$}
\address{$^1$ Research Group in Electronic, Biomedical and Telecommunication Engineering,
University of Castilla-La Mancha, Spain.}
\address{$^2$ BioMIT.org, Electronic Engineering Department, Universitat Politecnica de Valencia, Spain.}
\ead{\mailto{manuel.garcia@uclm.es}}


\begin{abstract}
\emph{Objective:} {The electrocardiogram (ECG) is nowadays the most widely used recording for diagnosis of cardiac disorders, including the most common supraventricular arrhythmia such as atrial fibrillation (AF).} However, different types of electrical disturbances, in which the power-line interference (PLI) is a major problem, can mask and distort the original ECG morphology. This is a significant issue in the context of AF, because accurate characterization of fibrillatory waves ($f$-waves) is still required to improve current knowledge about its mechanisms. This work introduces a new algorithm able to reduce high levels of PLI and preserve, simultaneously, the original ECG morphology. \emph{Approach:} The method is based on Stationary Wavelet Transform shrinking and makes use of a new thresholding function designed to work successfully in a wide variety of scenarios. {In fact, it has been validated on a general context with 48 ECG signals obtained from pathological and non-pathological conditions, as well as on a particular context of AF, where 380 synthesized and 20 long-term real ECG signals were analyzed.} \emph{Main results:} {In both situations the algorithm has reported a notably better performance than other common methods designed for the same purpose.} Moreover, its performance has proven to be optimal for dealing with ECG recordings affected by AF, since $f$-waves remained almost intact after removing very high noise levels. \emph{Significance:} The proposed algorithm may facilitate a reliable characterization of the $f$-waves, preventing them from not being masked by the PLI nor distorted by an unsuitable filtering applied to the ECG recording with AF.
\end{abstract}

\submitto{\PM}
\maketitle

\section{Introduction} \label{sec:intro}


Cardiovascular disease  is, in recent years, the leading global cause of death, representing 31\% of the mortality in the world. Whereas this disease was related to 17.3 million deaths per year in 2013, it is expected to reach 23.6 million by 2030, according to the World Health Organization~\cite{Anonymous:2017ug}. It includes conditions such as stroke, coronary heart disease, cardiomyopathy, heart failure and other cardiac disorders~\cite{Mendis:2011vd}. Among the latter, several heart rhythm abnormalities that are included under the term cardiac arrhythmias need to be highlighted, specially atrial fibrillation (AF). This is nowadays the most common arrhythmia, affecting about 33.5 million people worldwide~\cite{Chugh:2014kha}. {Despite its high prevalence, the mechanisms triggering and maintaining AF are not completely known, thus making its diagnosis and treatment poorly efficient~\cite{Schotten2016}. Hence, precise characterization of the electrocardiogram (ECG) recording plays a key role to improve current knowledge about this arrhythmia~\cite{Kotecha:2018il}.}

Unfortunately, the potentials recorded on the ECG are typically of only a few millivolts, thus different types of electrical disturbances can mask and distort the morphological features of this waveform, hindering the proper diagnosis of cardiac pathologies or even misleading the cardiologist or corrupting the proper processing function of automated systems~\cite{Poungponsri:2013exa}. Accordingly, the study of supraventricular arrhythmias, such as AF, needs the extraction of parameters from the interval between the T wave and the QRS complex (TQ Interval), where the atrial activity is manifested as multiple and random fibrillatory waves ($f$-waves), which are variable in shape and timing ~\cite{Garcia:2016km}. Taking into account their reduced amplitude with respect to the ventricular activity, a low signal-to-noise ratio can affect significantly to the atrial activity as has been reported by several previous studies~\cite{Larburu:2011wd}.

The electrical disturbances affecting the ECG can be caused by either biological or environmental sources~\cite{Asgari:2017di}. The first group includes motion artifacts, baseline wander and electromyographic noise. The environmental causes include power-line interference (PLI), and instrumentation noise. Although all of them can alter the ECG morphology, the PLI has special relevance, becoming one major common source of ECG interference. 

During ECG acquisition, nearby electronic devices and mains wiring connected to the power line will produce high amplitude interferent signals in the human body~\cite{Nagel:2000wg}. In addition, despite the high common mode rejection ratio of current biopotential amplifiers, their performance is mainly dependent on the high amount of electromagnetic noise that the human body picks up when considered as a volume conductor~\cite{Geselowitz1989IEEE}. Small differences in the skin-electrode impedance between electrodes also increase the common mode signal, thus providing a higher amplitude source of PLI~\cite{Levkov:2005kea}. In order to reduce the presence of PLI during ECG acquisition, different techniques have been proposed, including the use of batteries, analog filters, active electrodes, twisted input leads and different shielding methods~\cite{Jiang:2018fl}. But, despite these precautions, significant residual interferences remain, thus requiring signal pre-processing to improve later analyses.

A set of algorithms including different filtering stages are generally provided to attenuate or eliminate specific frequencies from the input signal, such as the PLI, as has been introduced by different studies~\cite{Bortolan:2015bj,Kora:2017kf,Mjahad:2015dj,Henriques:2015ua,Tayel:2018ju}. However, these filtering stages can alter important ECG features. For instance, PLI removal may result in changes in the QRS complex amplitude and slope, thus turning diagnosis of a variety of cardiac pathologies difficult~\cite{Maroofi:2015ta}. But also, PLI may affect to the information included in the TQ interval, which is essential for the study of AF and other supraventricular arrhythmias. To this respect, most papers published in recent years related to the study of AF from the surface ECG focus their effort on analyzing the TQ interval, either alone~\cite{Garcia:2016km,Rodenas_2015_ETP,Ladavich:2015fy}  or in combination with the study of an irregular ventricular rhythm~\cite{Rodenas:2017it,Petrenas:2015ik,Purerfellner:2014cc}.

In order to maintain signal fidelity and being able to perform later reliable analysis from the filtered ECG, different standardization organizations have proposed a maximum allowed noise value. The IEC 60601-2-25:2011 standard~\cite{Commision:2011uy} of the International Electrotechnical Committee establishes that the noise level must not exceed 30~$\mu V$ peak-to-valley in comparison with the original signal. Hence, any PLI denoising should remove the interference, but keeping the ECG signal practically intact. To this end, different methods have been proposed, the most conventional being based on a fixed frequency notch filter. Infinite Impulse Response (IIR) Butterworth and Chebyshev type I band-stop filters~\cite{Gupta:2013tv} or Butterworth and elliptic notch filters ~\cite{Chavan:2008tv,Narwaria:2011ti} have been used in this regard. The use of linear phase Finite Impulse Response (FIR) filters has been also applied by other authors, in combination with a variety of algorithms~\cite{Patel:2012tr} or by means of different windowing methods~\cite{Ahmad:2015uj}. Despite its simplicity, this type of filtering has the main disadvantage of performing as a function of the stability of the frequency to be eliminated. In addition, given that the ECG is a non-stationary signal, this sort of filtering often cause distortion of its frequency spectrum~\cite{Thalkar:2013ww}.

To overcome this problem, adaptive filtering has been employed in different works due to its ability to adjust the filter coefficients in real time according to the signal under processing, thus tracking most dynamic variations in the ECG. More precisely, Least Mean Square and Recursive Least Square algorithms, which operate by recursively adjusting the filter parameters, have been extensively used due to their low computational complexity~\cite{Manivel:2015tf,Manosueb:2014bs,Biswas:2014fh,Sushmitha:2014vk,Mursi:2015tq,Razzaq:2013bo,Tayel:2018ju}. Although these adaptive filters perform better than conventional fixed-bandwidth ones for time-varying signals, they still present the main shortcoming of requiring an external reference signal, which is not always easy to obtain~\cite{Sharma:2016ez}. Moreover, they often give rise to gradient noise, slow convergence rate~\cite{Zivanovic:2013cv} and report more distortion than other filtering methods {at the beginning of the adaptation procedure~\cite{Almahamdy:2014iv} and just after QRS complexes~\cite{Warmerdam:2017ki}.}

Some other approaches developed for PLI reduction include Empirical Mode Decomposition~\cite{Suchetha:2017fh,Rakshit:2018ex}, Kalman smoother ~\cite{Warmerdam:2017ki}, sparse decomposition~\cite{Zhu:2017ix}, recurrent neural networks~\cite{Yue:2017kq} and others based on a combination of several methodologies~\cite{Poungponsri:2013exa}. Among the different algorithms involved, wavelet transform (WT) has become a powerful tool for ECG processing due to its non-stationary nature. Thus, discrete WT (DWT) has been widely applied to ECG analysis in general and PLI denoising in particular by thresholding wavelet coefficients~\cite{Poungponsri:2013exa,Castillo:2013gba,Han:2016fsa,Poornachandra:2008ia}. Although two well-known approaches, such as soft and hard thresholding functions, have been traditionally employed for DWT-based denoising, they have shown some shortcomings~\cite{Donoho:1995jp}. Moreover, some efforts have also been made for the development of new thresholding functions able to deal with different kinds of noise in the ECG~\cite{Han:2016fsa}, but convincing results are yet to come. {Within this context, the present study introduces a novel wavelet-based denoising algorithm featured by an improved trade-off between PLI reduction and original ECG morphology preservation. }   
\par
{The method has been thoroughly validated on a broad variety of ECG recordings as well as on several realistic scenarios of PLI. On the one hand, ECG signals from healthy subjects and patients with different cardiac arrhythmias were firstly considered for a general validation of the algorithm. For this purpose, well-known and freely available databases were used. On the other hand, given the need of accurately featuring $f$-waves in the context of AF, synthesized and real ECG recordings obtained from patients with this arrhythmia were also analyzed. Moreover, in order to clearly assess the method's impact on the atrial and ventricular waves captured by the ECG, the TQ and QRST intervals were separately studied. To the best of our knowledge, this kind of evaluation for a PLI denoising algorithm is introduced for the first time in the present work. Finally, it should be noted that, in contrast to most previous studies where a sinusoidal signal of 50 Hz was only used as PLI, harmonic and interharmonic components of that dominant frequency were also considered. As well, two realistic scenarios showing large time-varying amplitude and frequency fluctuations in the PLI were studied. }

The remainder of the manuscript is organized as follows. {Section\S~\ref{sec:materials} describes the synthesized and real ECG signals used for validating the proposed method, which is introduced in Section~\S\ref{sec:methods}. In this part, other denoising algorithms included for comparison are also briefly described, along with the parameters used for their performance assessment.} Section~\S\ref{sec:results} summarizes the obtained results, which will be next discussed in Section~\S\ref{sec:discussion}. Finally, Section~\S\ref{sec:conclusions} presents the concluding remarks.

\section{{Materials}}\label{sec:materials}
\subsection{{Databases for general validation}}
{Most of previous works have used the MIT-BIH Arrhythmia database for validating the introduced denoising techniques~\cite{Poungponsri:2013exa,Manivel:2015tf,Sharma:2016ez,Han:2016fsa,Zhu:2017ix,Suchetha:2017fh,Rakshit:2018ex}. This database presents a balanced set of ECG recordings obtained from healthy subjects as well as patients suffering from complex ventricular, junctional and supraventricular arrhythmias and conduction abnormalities, thus drawing a general context where no special attention is paid to any particular cardiac pathology. More details can be found in~\cite{Moody2001}, but it should be noted that the dataset consists of 48 2-lead ECG recordings with a duration of 30 minutes. Although the signals were acquired with a sampling rate of 360~Hz, they were resampled at 1000~Hz.
}

\subsection{{Databases for particular validation on AF}}
For a thorough validation of the proposed denoising algorithm on the specific context of AF, a set of synthesized ECG recordings were firstly generated making use of previously published models. Briefly, the McSharry et al.'s method~\cite{McSharry:2003gy} was used to synthesize the ventricular activity with a sampling rate of 1000 Hz. The atrial activity was synthesized by using the model proposed by Stridh and Sörnmo~\cite{Stridh:2001jy}. Thus, $f$-waves were obtained as the sum of M harmonics with a sawtooth morphology and a frequency modulated around the fundamental one. Given that previous works have proven that the $f$-waves typical fundamental frequency ranges approximately from 3 to 9 Hz~\cite{Stridh:2001jy}, this parameter was randomly selected according to a normal distribution with mean and standard deviation of 6 and 1.5 Hz, respectively. The duration of each synthesized ECG signal, which will be referred to as $x(n)$ from now on, was 60 seconds.
\par
The use of synthesized ECG signals allowed us to control specific AF features that could influence denoising performance. To this respect, given that the TQ interval varies as a function of the heart rate, sets of 20 signals were synthesized for heart rates of 60, 80, 100, 120, 140, 160 and 180 beats per minute (bpm). Initially, no irregularity in RR intervals was considered to corroborate denoising ability to work under regular ventricular rhythm conditions. {This situation is common in presence of atrioventricular blocks, in ventricular or atrioventricular junctional tachycardias and in AF patients using pacemakers~\cite{Butta2013,Said2014}}. Nonetheless, in a second experiment new sets of 20 signals were generated with a heart rate selected according to a normal distribution with mean and standard deviation of 100 and 10~bpm, respectively, and {maximum variability between RR intervals of 0, 5, 10, 15, 20 and 25\% of the heart rate}. Note that in both tests the $f$-waves amplitude was established to its typical value of 75~$\mu$V~\cite{Stridh:2001jy}. Nonetheless, in a final experiment other sets of 20 signals were generated with a regular heart rate of 80~bpm and $f$-waves amplitudes of 15, 30, 60, 90, 120 and 150~$\mu$V. {A example of a typical synthesized ECG signal is shown in Figure~\ref{fig:figure1}(b).}   
\par
Finally, to assess the proposed PLI denoising ability on real AF episodes, the freely available MIT-BIH AF database was also used~\cite{Goldberger:2000up}. Although the dataset contains~23 fully annotated~10 hour-length ECG recordings obtained mainly from paroxysmal AF patients, only AF episodes (291 episodes, 474,670 beats and about 110 hours) were analyzed. In this case, the signals were also resampled at 1000~Hz in order to work with a common sampling rate in the study.

\subsection{Synthesis of the PLI}\label{sec:synth}
To obtain noisy signals, ECG recordings were corrupted by a simulated PLI. In Europe the standard EN50160 defines the main characteristics in public power distribution systems~\cite{Committee:1999wf}. Briefly, the power supply frequency is mainly set at~50~Hz with a maximum variation of~$\pm$1\%. Its normalized voltage is defined at 230~V with a maximum fluctuation about~$\pm$10\%. The signal can also present harmonic and interharmonic components; thus, their power being limited to 2, 5, 1 and~6\% for the first four multiples of 50~Hz, respectively, and to~0.2\% for inter-harmonics~\cite{Committee:1999wf}. To mimic this interference as realistic as possible, all these aspects were considered, thus including random amplitude and frequency variations within the described limits for the main frequency of 50~Hz, as well as for its first four harmonics. Some low-amplitude interharmonics were also considered through a frequency modulation of each frequency component with a maximum deviation of 0.5~Hz. The resulting interference was named common PLI and was added to the ECG recordings for obtaining noisy signals (referred to as $\tilde{x}(n)$) with input signal-to-noise (SNR$_{\mathrm{in}}$) ratios of 15, 10, 5, 0, $-$5 and $-$10 dB.

However, previous works have shown that the PLI observed in biomedical recordings, including ECG signals, is prone to exhibit variations beyond the limits established by the standard EN50160~\cite{Levkov:2005kea,Zivanovic:2013cv}. This aspect mainly owes its existence to the unstable and nonlinear behavior of some electronic equipments, such as transformers, lamps, etc.~\cite{Costa:2009hu}. Thus, to consider this real context, the proposed algorithm was also validated on some extended conditions of PLI. On the one hand, a sudden sinusoidal amplitude variation in the PLI was simulated by abruptly reducing the SNR$_{\mathrm{in}}$ from infinite to a specific value in the tenth second of the signal~\cite{Warmerdam:2017ki}. Moreover, sinusoidal variation of the PLI amplitude with a low-frequency between~0.5 and~2 Hz was also considered for the remaining~50 seconds~\cite{Warmerdam:2017ki}. On the other hand, to study the influence of high frequency deviations of the theoretical PLI frequency, deviations of $\pm$3~Hz were introduced in the main component of~50 Hz and its harmonics~\cite{Zivanovic:2013cv}. As before, SNR$_{\mathrm{in}}$ values of 15, 10, 5, 0, $-$5 and $-$10~dB were considered for the two experiments.


\section{Methods}\label{sec:methods}

\subsection{Stationary wavelet shrinking for PLI denoising}\label{sec:shirnk}
Given its ability to treat with non-stationary signals, the WT is a powerful tool that has been applied to a broad variety of research fields~\cite{Addison:2017vl}. In contrast to the traditional Fourier transform, the WT provides a time-frequency analysis able to detect local, transient or intermittent aperiodicites in the signal. Indeed, it is a linear transform, which decomposes a signal at several levels {by computing its convolution with scaled and shifted versions of a mother wavelet. Detailed mathematical expressions related to this tool can be found in \cite{Addison:2017vl}}. For practical applications, the DWT has been immensely used because it can be easily implemented by a filter bank. This methodology is able to provide useful time-frequency information about the signal, whereas only requiring reasonable computation time. However, the DWT does not preserve translation invariance, since sub-sampling is applied to the filtered signal to obtain each scale. This aspect can sometimes cause problems of repeatability and robustness, especially when too short signals are analyzed~\cite{Asgari:2015iqa}. Hence, in the present work a time-invariant alternative of the DWT, named stationary WT (SWT), was used. This tool makes use of recursively dilated filters, instead of subsampling, to halve the bandwidth from one scale to another~\cite{Asgari:2015iqa}.
 
Motivated by the idea that WT is able to provide a sparse representation of useful data and nuisance interferences, this tool has been used to reduce noise in a broad variety of signals. In fact, if a suitable orthogonal mother wavelet is chosen, the signal decomposition can separate its main profile from noise and other nuisance interferences~\cite{LijunXuandYong:2004wc}. More precisely, whereas the most relevant details of the signal are represented by high-amplitude wavelet coefficients, noise is mainly associated with low-amplitude coefficients. Then, the nuisance interference can be mostly reduced, without modifying substantially the original signal morphology, by zeroing that coefficients~\cite{LijunXuandYong:2004wc}. This approach has proven to be highly efficient to reduce white and gaussian high-frequency noise in a variety of biomedical recordings, including the ECG~\cite{ElBcharri:2017jd}. Nonetheless, a few works have also shown its ability to remove sinusoidal perturbations from this signal~\cite{Poornachandra:2008ia,LijunXuandYong:2004wc}. In view of this context, a SWT-based method is proposed in this manuscript aimed at removing PLI {in ECG recordings obtained from a wide variety of pathological and non-pathological conditions}. 

\begin{figure*}[tp!]
\centering
\includegraphics[width=\textwidth,keepaspectratio]{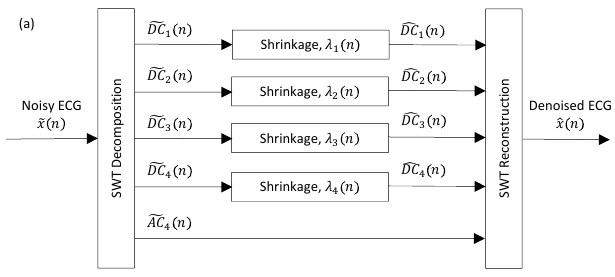}\\
\includegraphics[width=\textwidth,keepaspectratio]{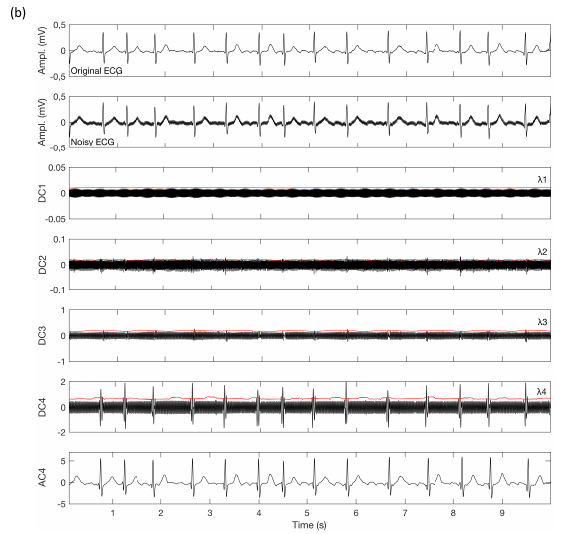}
\caption{(a) Block diagram displaying the main steps for the proposed SWT-based denoising algorithm. (b) Typical wavelet coefficients yielded by decomposition and the corresponding thresholds $\lambda_j(n)$.}\label{fig:figure1}
\end{figure*}

A block diagram summarizing the proposed strategy is displayed in Figure~\ref{fig:figure1}(a). As can be seen, the noisy ECG recording is firstly decomposed into four wavelet levels and detail coefficients for scales 1, 2, 3 and 4 (referred to as $\widetilde{DC}_1(n)$, $\widetilde{DC}_2(n)$, $\widetilde{DC}_3(n)$ and $\widetilde{DC}_4(n)$) are then thresholded. Given that ECG signals were acquired or resampled at 1000~Hz, the main PLI component is located in the detail coefficient of scale 4 (32.25$-$62.5~Hz) and its harmonics in the detail coefficients of scales 3 (62.5$-$125~Hz), 2 (125$-$250~Hz) and 1 (250$-$500~Hz), therefore, the approximation coefficient for scale 4 (0$-$31,25~Hz) does not require to be thresholded. In fact, the ECG bandwidth is mainly confined in low frequencies, below 30 Hz~\cite{Sornmo:2005wk}. As a final step, the denoised ECG signal is reconstructed by applying inverse SWT both to the shrunk detail coefficients for scales 1, 2, 3 and 4 as well as to the unmodified approximation coefficient for scale 4.

An essential aspect for successful PLI suppression is the selection of a proper threshold within each scale, because this cut-off value acts as an oracle discerning between relevant and irrelevant wavelet coefficients. Although different alternatives are well-known to estimate an appropriate threshold for white noise elimination, such as the universal approximation, the Stein's unbiased risk estimation, the minimax approximation, etc.~\cite{ElBcharri:2017jd}, no much information is provided to this respect for the case of a sinusoidal interference. Nonetheless, according to few previous works~\cite{Poornachandra:2008ia,LijunXuandYong:2004wc}, an adaptive threshold for each scale was computed by discarding high-amplitude information in wavelet coefficients, such as Figure~\ref{fig:figure1}(b) shows for a common example. Precisely, for the detail coefficient $\widetilde{DC}_j(n)$ the threshold $\lambda_j(n)$ was obtained as the result of applying a moving median filter for a window of 200 ms to the absolute value of the wavelet coefficients. 

Finally, it is worth noting that the way in which wavelet coefficients are thresholded has also a relevant impact on noise suppression~\cite{ElBcharri:2017jd}. As commented in Section~\S\ref{sec:intro}, the well-established hard and soft thresholding functions present some limitations~\cite{Donoho:1995jp}. In fact, the hard thresholding is highly sensitive to small changes in data, thus sometimes generating oscillations in the reconstructed signal~\cite{Donoho:1995jp}. The soft alternative overcomes this problem, but a bias is introduced in large wavelet coefficients, thus causing subtly blurred edges in the reconstructed signal~\cite{ElBcharri:2017jd}. Although other proposals have provided an intermediate behavior between these two approaches~\cite{ElBcharri:2017jd}, a combination of them is proposed to exploit the main characteristics of the ECG morphology. The main idea is to reduce as much noise as possible in those intervals between successive QRS complexes, but simultaneously preserving unaltered these impulsive waveforms. Hence, whereas a hard thresholding has been applied in areas associated to QRS complexes, a soft shrinkage has been used for the remaining regions. To discern easily between both cases, QRS complexes were experimentally identified by using 1.5 times the threshold $\lambda_j(n)$. Thus, the final shrinkage approach was defined as
\begin{equation}
\widehat{DC_j}(n) = \left\{\begin{array}{ll}
0, &\textrm{if } |\widetilde{DC_j}(n)| \leq \lambda_j(n),\\
\widetilde{DC_j}(n)-\lambda_j(n), &\textrm{if } \lambda_j(n)<\widetilde{DC_j}(n) \leq 1.5\cdot \lambda_j(n), \\
\widetilde{DC_j}(n), &\textrm{if } \widetilde{DC_j}(n) > 1.5\cdot\lambda_j(n),\\
\widetilde{DC_j}(n)+\lambda_j(n), &\textrm{if } -1.5\cdot\lambda_j(n) < \widetilde{DC_j}(n) \leq -\lambda_j(n),\\
\widetilde{DC_j}(n), &\textrm{if } \widetilde{DC_j}(n) < -1.5\cdot\lambda_j(n).
\end{array} \right.
\end{equation}  

\subsection{Reference methods for PLI denoising}\label{sub:others}
With the aim to compare the proposed method's performance, two representative notch filtering approaches have also been studied. On the one hand, a fixed-bandwidth and 2nd-order Butterworth filter, centered at 50~Hz and with a band-stop of $\pm$1~Hz, was implemented {according to the indications found in  \cite{Gupta:2013tv}}. On the other hand, {the adaptive notch filtering proposed by Costa \& Tavares was also developed with a convergence step-size of 0.1~\cite{Costa:2009hu}.}
\par
{Additionally, three common thresholding methods for wavelet-based denoising have also been considered for comparison. Thus, popular hard and soft thresholding functions were analyzed, along with an alternative proposed to overcome their drawbacks, i.e. the hyperbolic thresholding~\cite{Poornachandra:2008ia}. The threshold for each scale was computed making use of the well-known minimax estimation, since previous works have proven its efficiency for dealing with ECG recordings~\cite{Castillo:2013gba}.  
}  


\subsection{Performance assessment}\label{sub:perform}
The PLI reduction and the waveform integrity preservation ability of the proposed denoising algorithms were validated by computing the output signal-to-noise ratio (SNR$_{\mathrm{out}}$) and an adaptive signed correlation index (ASCI) between the denoised signal $\widehat{x}(n)$ and the original clean $x(n)$. The SNR$_{\mathrm{out}}$ was defined as 
\begin{equation}
\mathrm{SNR}_{\mathrm{out}}=P_{\widehat{x}}/P_r,
\end{equation}
where $P_{\hat{x}}$ is the power of the denoised output signal $\widehat{x}(n)$, and $P_r$ is the power of the remaining PLI~$(r=\widehat{x}-x)$. Considering $N$ sample-length signals, the ASCI was defined as
\begin{equation}
\mathrm{ASCI}\big(x(n),\hat{x}(n)\big)=\frac{1}{N}\sum_{k=1}^{N}x(k)\otimes\hat{x}(k), 
\end{equation}
where $\otimes$ denotes the signed product of two dicotomized scalars as
\begin{equation}
x(n)\otimes\hat{x}(n)=\left\{
\begin{array}{rl}
1 & \mathrm{if}~|x(n)-\hat{x}(n)| \leq \beta,\\
-1 & \mathrm{if}~|x(n)-\hat{x}(n)| > \beta.
\end{array} \right.
\end{equation}
The threshold $\beta$ was experimentally selected as 5\% of the standard deviation of $x(n)$. It is interesting to note that, unlike the well-established Pearson's correlation index, the ASCI is able to consider also amplitude differences between two signals for their morphological comparison~\cite{Lian:2010kj}.
\par
{In addition to these two global parameters, the ASCI was also separately computed for TQ and QRST intervals. In this way, particular information about the atrial and ventricular components captured by the ECG was accurately obtained.} Both intervals were segmented as in previous works~\cite{Garcia:2016km,Rodenas_2015_ETP}. In short, the TQ interval was selected as a window of varying size preceding a reference point for each beat that was placed 50~ms before the R-peak. Because this interval length is highly variable with the heart rate~\cite{Fossa:2010wp}, it was adaptively selected as a quarter of the mean RR interval for the last five beats~\cite{Garcia:2016km}. {The remaining interval for each beat was then considered as the QRST segment}.


\section{Results}\label{sec:results}
As most WT-based algorithms~\cite{Addison:2017vl}, the performance in PLI reduction of the proposed SWT-based method was also dependent on the selected mother wavelet. Thus, because no systematic rules can be found in the literature to choose that function, an exploratory approach based on testing different functions was adopted. All the functions from Haar, Daubechies, Coiflet, Biorthogonal, Reverse Biorthogonal and Symlet wavelet families were analyzed. Although no significant differences were observed, the best outcomes were obtained from the sixth-order Daubechies function. Hence, this wavelet mother was chosen, and its main results will be presented as it follows.

\subsection{{Validation of the denoising algorithms on a general context}}
\begin{figure*}[tp!]
\centering
\includegraphics[width=\textwidth,keepaspectratio]{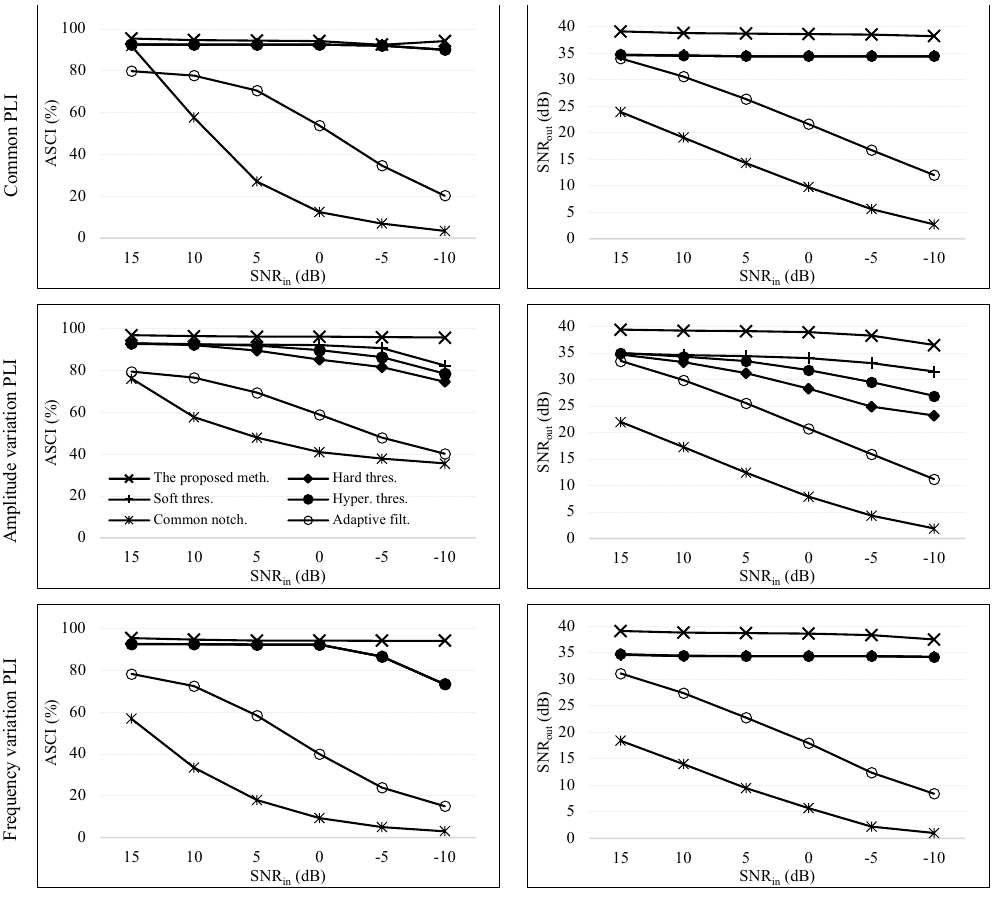}
\caption{Waveform integrity preservation, ASCI (\%), and PLI reduction, $\mathrm{SNR_{out}}$ (dB), from the ECG recordings collected by the MIT-BIH Arrhythmia database, provoked by the analyzed methods after denoising of a common PLI, a PLI with time-varying amplitude, and a PLI with frequency deviations between $\pm$3~Hz.}\label{fig:arrhytmia}
\end{figure*}

{Figure~\ref{fig:arrhytmia} shows the denoising methods' performance in terms of the global parameters ASCI and $\mathrm{SNR_{out}}$ for the three considered scenarios of PLI. As can be observed, the proposed SWT-based algorithm presented a stable behavior for every condition of PLI as well as for every value of $\mathrm{SNR_{in}}$, always reporting values of ASCI and $\mathrm{SNR_{out}}$ greater than 95\% and 37~dB, respectively. The other three wavelet-based denoising methods only displayed a similar stable performance for the case of a common PLI, although slightly lower values of ASCI and $\mathrm{SNR_{out}}$ were noticed (about 92\% and 34~dB, respectively). Moreover, when amplitude and frequency variations were considered in the PLI, hard, soft and hyperbolic thresholding functions provided a notable worsening for values of $\mathrm{SNR_{in}}$ lower than 0~dB. Finally, it should be noted that the two notch filtering approaches always reached notably lower values of ASCI and $\mathrm{SNR_{out}}$ than the wavelet-based algorithms. Moreover, the performance of these filters was highly dependent on the noise level, such that values of ASCI and   $\mathrm{SNR_{out}}$ lower than 50\% and 25~dB were respectively noticed for rates of $\mathrm{SNR_{in}}$ lower than 0~dB. 
} 

\begin{figure*}[tp!]
\centering
\includegraphics[width=\textwidth,keepaspectratio]{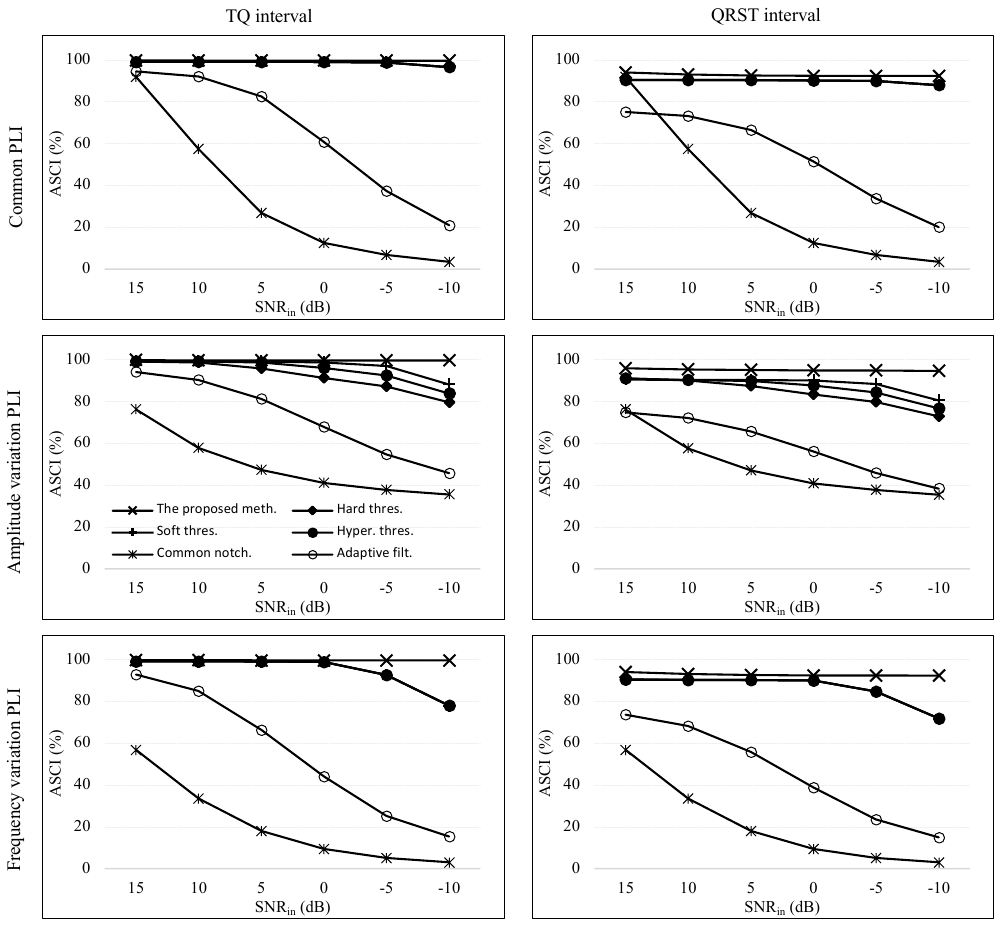}
\caption{Waveform integrity preservation, ASCI (\%), in TQ and QRST intervals from the ECG recordings collected by the MIT-BIH Arrhythmia database, provoked by the analyzed methods after denoising of a common PLI, a PLI with time-varying amplitude, and a PLI with frequency deviations between $\pm$3~Hz.}\label{fig:arrhytmia_TQ}
\end{figure*}

{The values of ASCI separately obtained for TQ and QRST intervals are shown in Figure~\ref{fig:arrhytmia_TQ}. In this case, all the tested denoising methods reported a behavior roughly similar to the global case (displayed in Figure~\ref{fig:arrhytmia}) both for TQ and QRST excerpts. Of note is only that, in general, values of ASCI were notably higher for TQ intervals than for QRST segments. For instance, wheres values about 100\% were reported by the proposed SWT-based algorithm for TQ intervals, about 90\% were noticed for QRST segments.   
}

\subsection{{Validation of the denoising algorithms on the context of AF}}
\begin{figure*}[tp!]
\centering
\includegraphics[width=\textwidth,keepaspectratio]{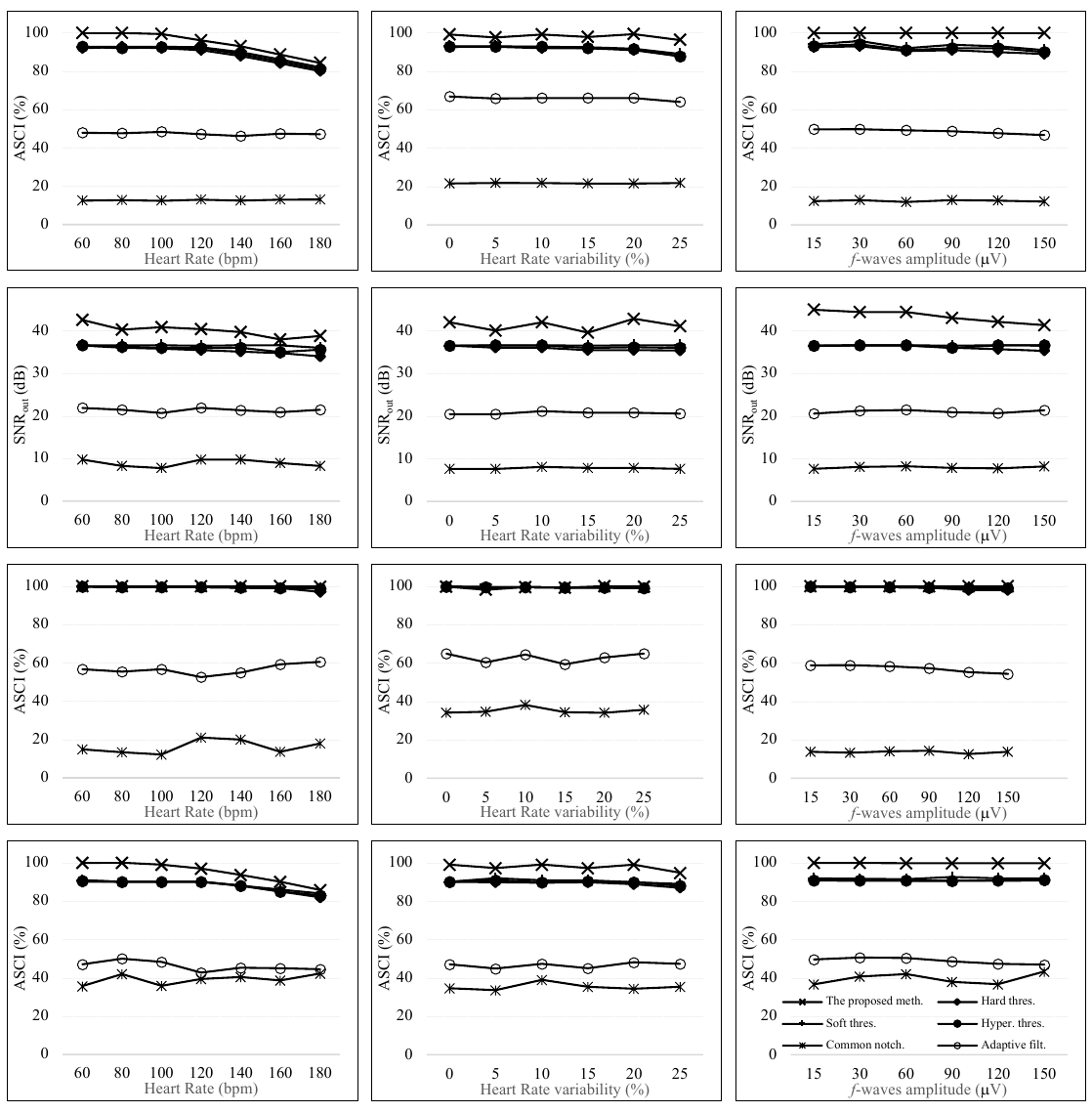}
\caption{Results obtained by applying the denoising methods to the ECG recordings synthesized with specific AF features, i.e., with different heart rates (plots in the first column), levels of irregularity in RR intervals (plots in the second column) and $f$-waves amplitudes (plots in the third column). The analyzed performance indices were ASCI globally computed from the whole original and denoised signals (plots in the first row) and separately obtained for TQ intervals (plots in the third row) and QRST segments (plots in the fourth row) and $\mathrm{SNR_{out}}$ (plots in the second row).   }\label{fig:synthesized_ECG}
\end{figure*}

The results obtained by applying the denoising algorithms to the ECG recordings synthesized with specific AF features are shown in Figure~\ref{fig:synthesized_ECG}. A common PLI with a $\mathrm{SNR_{in}}$ of 0~dB was only considered in this experiment, because no significant variations were observed for the remaining scenarios and noise levels. It is interesting to note that the two notch filters presented a highly stable behavior in terms of ASCI (globally computed for the whole original and denoisied signals and separately obtained for the TQ and QRST intervals) and $\mathrm{SNR_{out}}$ independently on the heart rate, the irregularity in RR intervals and the $f$-waves amplitude. A similar behavior was also noticed for the wavelet-based algorithms, except for heart rates higher than 140 bpm where the performance indices showed a slight decrease. Anyway, these methods always presented higher values of ASCI and $\mathrm{SNR_{out}}$ than both fixed-bandwidth and adaptive filtering approaches. Thus, the proposed SWT-based denoising technique yielded the best outcomes, reporting values of ASCI and $\mathrm{SNR_{out}}$ close to 100\% and 38~dB, respectively, for most tests.

\begin{figure*}[tp!]
\centering
\includegraphics[width=\textwidth,keepaspectratio]{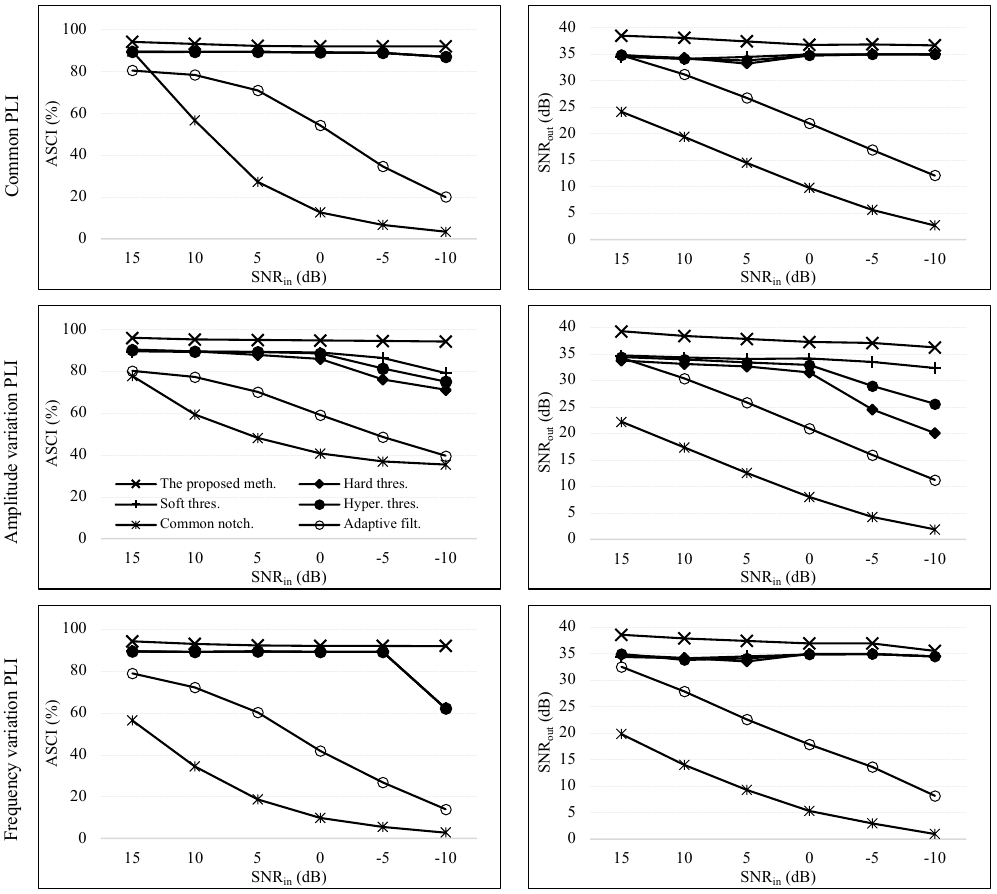}
\caption{Waveform integrity preservation, ASCI (\%), and PLI reduction, $\mathrm{SNR_{out}}$ (dB), from the real AF episodes collected from the MIT-BIH AF database, provoked by the analyzed methods after denoising of a common PLI, a PLI with time-varying amplitude, and a PLI with frequency deviations between $\pm$3~Hz.}\label{fig:af}
\end{figure*}

\begin{figure*}[tp!]
\centering
\includegraphics[width=\textwidth,keepaspectratio]{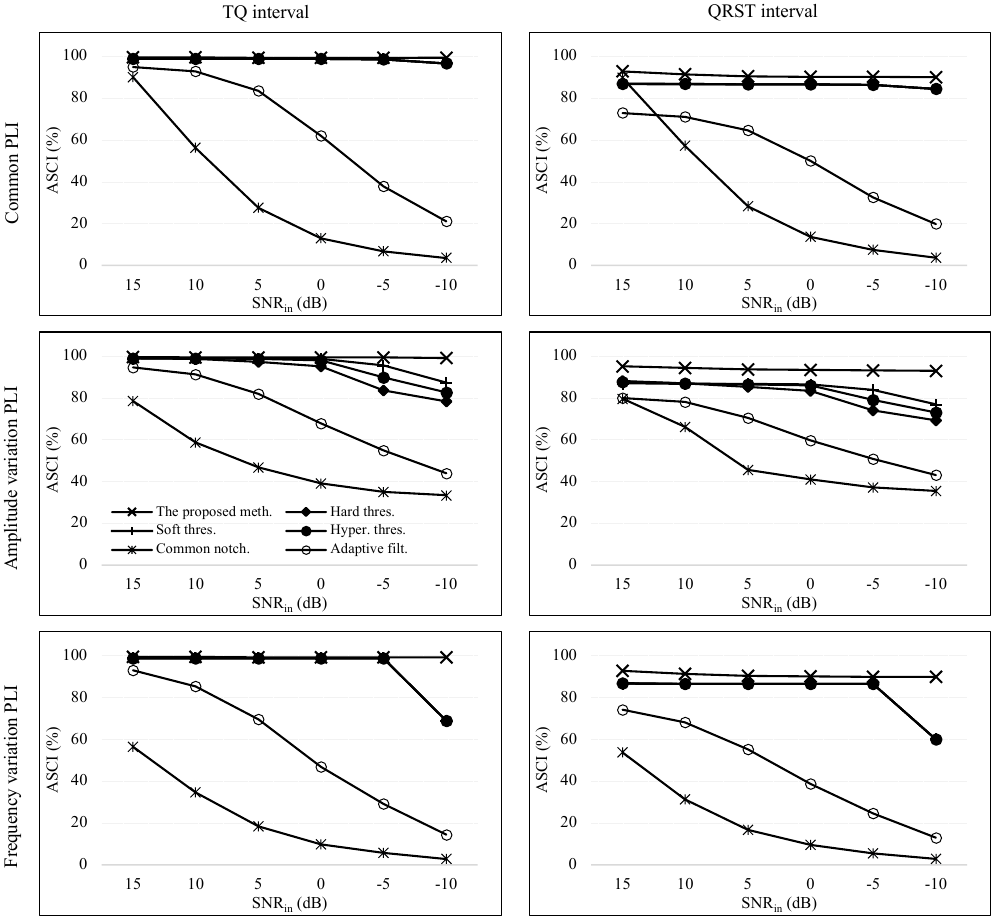}
\caption{Waveform integrity preservation, ASCI (\%), in TQ and QRST intervals from the real AF episodes collected from the MIT-BIH AF database, provoked by the analyzed methods after denoising of a common PLI, a PLI with time-varying amplitude, and a PLI with frequency deviations between $\pm$3~Hz.}\label{fig:af_TQ}
\end{figure*}

{Regarding the denoising of real AF episodes extracted from the MIT-BIH AF database, Figure~\ref{fig:af} shows evolution of the global parameters ASCI and $\mathrm{SNR_{out}}$ as a function of the noise level. Compared with the results presented in Figure~\ref{fig:arrhytmia} for the MIT-BIH Arrhythmia database, no relevant differences can be highlighted for any algorithm and any condition of PLI. In the same way, the values of ASCI separately computed for TQ and QRST intervals, displayed in Figure~\ref{fig:af_TQ}, were also quite similar to those obtained for the general context (see Figure~\ref{fig:arrhytmia_TQ}). 
}

\begin{figure*}[p!]
\centering
\includegraphics[width=\textwidth,keepaspectratio]{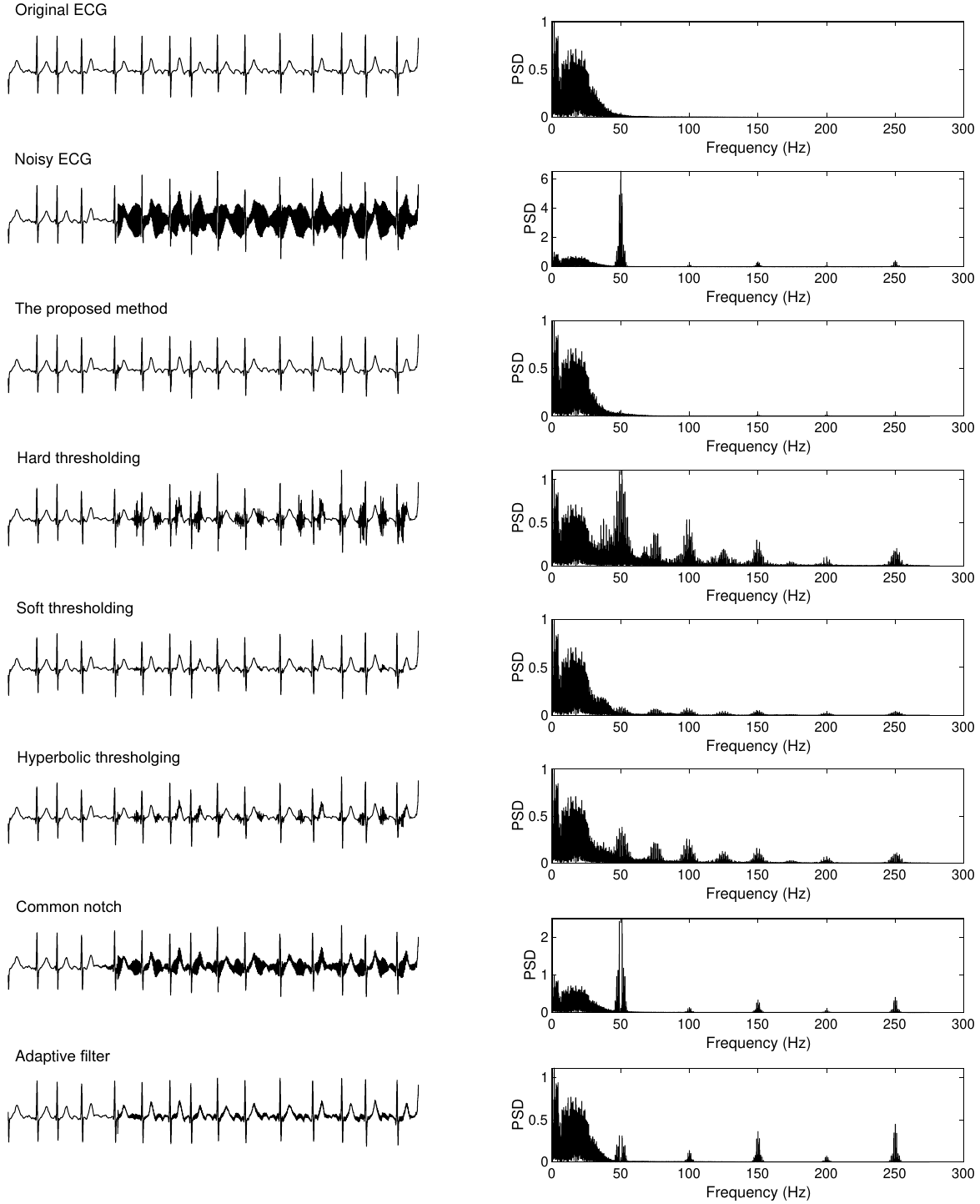}
\caption{Example of real ECGs displaying the denoising performance of the analyzed algorithms for a PLI with time-varying amplitude together with their spectral distributions. Note that a 10 second-length real AF recording with a $\mathrm{SNR_{in}}$ of 0 dB was analyzed.}\label{fig:example}
\end{figure*}

As a graphical summary, Figure~\ref{fig:example} displays the ECG signals obtained with all denoising algorithms from a typical real AF episode corrupted with a PLI presenting a time-varying amplitude and a $\mathrm{SNR_{in}}$ of 0~dB. Note that, for illustrative purposes, spectral distributions from all the signals are also presented.

\section{Discussion}\label{sec:discussion} 
{To the best of our knowledge, in the present work several PLI denoising algorithms have been validated for the first time on a broad variety of ECG recordings, thus paying attention both to a general context and to the particular situation of AF. Interestingly, the methods have reported a similar behavior for the two cases with respect to PLI reduction and ECG morphology preservation. In fact, the values of ASCI and $\mathrm{SNR_{out}}$ globally computed from the ECG signals collected by the MIT-BIH Arrhythmia and AF databases have provided nearly identical trends for all algorithms (compare Figures~\ref{fig:arrhytmia} and~\ref{fig:af}), thus suggesting the same impact on ECG recordings acquired from pathological and non-pathological conditions. Moreover, for a more accurate analysis about how morphology of atrial and ventricular waves captured by the ECG were preserved after denoising, this work has also been pioneer in separately assessing TQ and QRST intervals. Again, all tested denoising techniques have also provided a similar performance for the analyzed datasets (see Figures~\ref{fig:arrhytmia_TQ} and~\ref{fig:af_TQ}).         
}
\par
Some previous works have shown a relatively good behavior for PLI reduction by applying a fixed-bandwidth notch filtering to the overall ECG recording~\cite{Gupta:2013tv,Ahmad:2015uj}. Nevertheless, in a large amount of these studies, the method has not been exposed to tests aimed at verifying its performance when interferences of time-varying amplitude and/or frequency are considered. In fact, the results obtained in the present study have only reported a good performance of this type of filtering for the largest $\mathrm{SNR_{in}}$ values (15~dB) under conditions of a common PLI. Thus, even when this condition was maintained, the results worsened significantly for lower levels of $\mathrm{SNR_{in}}$, as shown in Figures~\ref{fig:arrhytmia} and~\ref{fig:af}. Consequently, this algorithm is unable to preserve the original ECG waveform morphology for reduced values of $\mathrm{SNR_{in}}$. {This finding has also been suggested by the separate analysis of TQ and QRST intervals, which has revealed no significant differences in the method's impact on the morphology of both segments.}
\par
According to some recent works~\cite{Zivanovic:2013cv,Rakshit:2018ex}, the obtained results have also reported a significant dependence on the performance of fixed-bandwidth notch filtering with a time-varying amplitude PLI and with a changing frequency PLI, leading to an overall decrease in the ASCI close to 40\% and 60\%, respectively, and close to 20 dB in $\mathrm{SNR_{out}}$ for both cases. These poor outcomes may be explained, on the one hand, because having fixed parameters this type of filtering cannot track the drift in the disturbing frequency and, on the other hand, because of the well-known ringing effect~\cite{Razzaq:2013bo}. This effect introduces distortion in regions where fast changes from higher to lower bandwidth are produced. Thus, notable distortion is observed in the ECG after QRS complexes, and will always be displayed, both for the largest values of $\mathrm{SNR_{in}}$ and even when PLI is not present in the signal to be filtered~\cite{Warmerdam:2017ki}. 
\par
Regarding the adaptive notch filtering, in general it has provided a better performance than the fixed-bandwidth notch alternative, thus showing a greater efficiency in the tracking of PLI fluctuations. To this respect, the fixed-bandwidth notch filter only performed better for the largest $\mathrm{SNR_{in}}$ of 15~dB and a common PLI (see Figures~\ref{fig:arrhytmia} and~\ref{fig:af}). For reduced $\mathrm{SNR_{in}}$ values, the behavior of the adaptive filter was better, always maintaining a constant differential of approximately 12~dB in $\mathrm{SNR_{out}}$ for any $\mathrm{SNR_{in}}$ value. Nonetheless, the ability of this method to preserve the original ECG morphology was still far from being optimal, showing ASCI values lower than 65\% for $\mathrm{SNR_{in}}$ rates lower than 0~dB. Similarly, the algorithm was also unable to maintain proper functionality when exposed to frequency and amplitude PLI variations. This poor behavior could be explained because the attenuation achieved by the filter at the frequency of the PLI depends on its capability of adaptation, which varies with the SNR~\cite{Verma:2015je}. Furthermore, the rapid adaptation rates that are required to track fluctuations in the PLI are unable to avoid the interference of the QRS complexes in the adaptive procedure. {This latter aspect also seems to be responsible for the notably different ASCI values computed for TQ and QRST intervals (see Figures~\ref{fig:arrhytmia_TQ} and~\ref{fig:af_TQ}). In fact, whereas a similar behavior was observed for both ECG segments, $f$-waves in the TQ intervals were notably less distorted than the QRST complexes, indeed observing improvements between 20 and 5\%. 
} 
\par
{In comparison with these two filtering approaches, all wavelet-based denoising methods revealed a significantly better trade-off between PLI reduction and ECG morphology preservation. They displayed a stable behavior for every noise level  and most conditions of PLI. In fact, global values of ASCI and $\mathrm{SNR_{out}}$ higher than 90\% and 35~dB were always observed for a common PLI and most $\mathrm{SNR_{in}}$ levels when amplitude and frequency variations were considered in the interference (see Figures~\ref{fig:arrhytmia} and~\ref{fig:af}). Nonetheless, despite this promising outcome for the traditional hard, soft and hyperbolic thresholding functions, the proposed denoising algorithm still reached a better performance. Thus, even in the case of a common PLI, where the lowest differences were noticed among wavelet-based methods, the proposed approach reported improvements about 5\% and 5~dB in global values of ASCI and $\mathrm{SNR_{out}}$, respectively. This outcome may be explained because the largest wavelet coefficients (i.e. those higher than $1.5\cdot\lambda_j(n)$) are preserved without modification through a hard thresholding, thus provoking less morphological alteration in the  QRS complexes. In fact, Figures~\ref{fig:arrhytmia_TQ} and~\ref{fig:af_TQ} show how all wavelet-based algorithms preserved largely intact the TQ intervals (ASCI values higher than 97\%), but the proposed denoising method reported improvements between 4 and 10\% in the QRST segments. 
}

{This last result also leads to another relevant finding. More precisely, the fact that all wavelet-based denoising methods marginally modified the TQ interval suggests that they are especially useful for PLI removal from ECG recordings affected by AF.} In fact, large preservation of $f$-waves is essential to conduct reliable studies on this arrhythmia, which are still imperative~\cite{Kotecha:2018il}. Nowadays, there exists a lack of tools capable of carrying out an accurate estimation of the arrhythmogenic substrate of every patient in AF, thus cardiologists have the most reliable information  on the arrhythmogenic patient status in order to choose the most appropriate personalized treatment. The development of new and more effective signal preprocessing tools, able to preserve as much as possible the original atrial activity information, could be helpful to identify worthwhile information able to guide the specialist in the choice of the best treatment.
\par
{Interestingly, all the wavelet-based denoising methods also reported a highly stable behavior when substantial variations were considered in specific features of AF (see Figure~\ref{fig:synthesized_ECG}).}  Thus, the outcomes from synthesized ECG recordings confirmed global values of ASCI close to 100\% and $\mathrm{SNR_{out}}$ near 40~dB, both when the algorithms were exposed to changes in heart rate variability and in $f$-waves amplitude. Only an appreciable decrease about 15\%  was observed in global values of ASCI for all methods when heart rates varied from 60 to 180~bpm. {However, this falling trend seems to mainly come from the increased morphological alteration caused by the algorithms on the QRST intervals for heart rates greater than 100~bpm. In fact, the ASCI values computed from the TQ segments remained above 97\% for every heart rate.}
\par
{Nonetheless, even in this particular context of AF, it is still worth noting that the proposed denoising algorithm presented a better performance than the remaining wavelet-based methods, specially when amplitude variations were considered in the PLI. Thus, Figures~\ref{fig:arrhytmia_TQ} and~\ref{fig:af_TQ} show how the common soft, hard and hyperbolic thresholding functions significantly reduced their ability to preserve the original ECG morphology in the TQ intervals for $\mathrm{SNR_{in}}$ values lower than 5~dB. On the contrary, the proposed algorithm maintained a stable behavior for every noise level. This outcome could be explained by the way in which the threshold was selected. Thus, whereas the used moving median filtering was able to track amplitude variations in the PLI, traditional alternatives for that purpose (such as the minimax estimation), only compute a single time-independent value for each wavelet scale, thus hampering removal of non-stationary interferences~\cite{Castillo:2013gba}.} However, this kind of time-varying PLI is often present in ECG recordings, because it results from changes in the electric/magnetic couplings between the source of interference and the human body. Hence, small alterations, like changes in the patient's position, can lead to significant fluctuations in the amplitude of this interference and its reduction was a important challenge for all denoising algorithms proposed in previous works~\cite{Warmerdam:2017ki}.
\par
{Finally, it should be noted that the proposed denoising algorithm was also tested on ECG recordings obtained with a modern acquiring device, such as the AliveCor one. In short, the sample set proposed for the Physionet/CinC Challenge 2017~\cite{Clifford2017} was analyzed. This dataset is not focused on ECG signals acquired from any pathological or non-pathological conditions, because it is formed by 270 excerpts lasting from 9 to 60 seconds and containing episodes of normal sinus rhythm (150), AF (50) and other unspecified rhythms (70). The yielded outcomes were closely similar to those displayed in Figures~\ref{fig:arrhytmia}, \ref{fig:arrhytmia_TQ}, \ref{fig:af} and~\ref{fig:af_TQ} and, therefore, they were not presented here. Nonetheless, they suggest that the proposed method can also work successfully with ECG recordings acquired with devices that will play a key role in diagnosis and long-term monitoring of many cardiac diseases in the near future~\cite{Lip2017}. 
}


\section{Conclusions}\label{sec:conclusions}
A new filtering methodology based on SWT shrinking has been proposed to reduce high levels of PLI in the ECG recording and, simultaneously, preserve its original morphology. {The method has been validated on a general context by analyzing ECG signals obtained from pathological and non-pathological conditions, as well as on the particular scenario of AF.} In both cases its performance has been notably better than common fixed-bandwidth and adaptive notch filters, { as well as than other three well-established wavelet-based denoising strategies.} {The separate study about how the morphology of TQ and QRST intervals is preserved after denoising has also revealed that the algorithm is optimal for the denoising of ECG signals affected by AF, because $f$-waves remained largely intact even for very high noise levels.} Additionally, the method has also shown to be immune not only to fluctuations in amplitude and frequency of the PLI, but also to changes in ECG features related to AF, such as the heart rate, its variability and the varying $f$-waves amplitude. As a consequence, this new methodology may facilitate the reliable characterization of the $f$-waves in AF, preventing them from not being masked by PLI nor distorted by an unsuitable filtering applied to the ECG signal.
 
\section*{Acknowledgements}
Research supported by the grants DPI2017--83952--C3 MINECO/AEI/FEDER, UE and SBPLY/17/180501/000411 from Junta de Comunidades de Castilla-La Mancha

\section*{Conflicts of Interest}
The authors declare no conflict of interest.

\section*{References}


\end{document}